\documentclass[twocolumn,showpacs,aps,prd]{revtex4-1}

\usepackage{graphicx}
\usepackage{dcolumn}
\usepackage{amsmath}
\usepackage{epsfig}
\usepackage{subfigure}
\usepackage{epstopdf}
\usepackage{multirow}
\usepackage{mathrsfs}
\usepackage[dvipdfmx, bookmarksnumbered, pdfstartview=FitH,colorlinks,urlcolor=blue, citecolor=blue,linkcolor=blue,] {hyperref}
\usepackage{lineno}
\input babarsym
\usepackage{ulem}
\usepackage{enumerate}
\usepackage{float}

\newcommand{\etap}{\eta^{\prime}}

\newcommand{\psip}{\psi(3686)}

\newcommand{\dz}{$D^{0}$~}

\newcommand{\jpsiphsp}{J/\psi\rightarrow \pi^{+} \pi^{-}\etap}

\newcommand{\jpsidimu}{J/\psi\rightarrow\mu^{+} \mu^{-}\etap}
\newcommand{\jpsirho}{J/\psi\rightarrow \rho \etap}

\newcommand{\jpsiomega}{J/\psi\rightarrow \omega \etap}
\newcommand{\psipphsp}{\psip\rightarrow\pi^{+} \pi^{-}\etap}
\newcommand{\psiprho}{\psip\rightarrow \rho \etap}

\newcommand*{\rom}[1]{\expandafter\@slowromancap\romannumeral #1@}
\def\deg{${}^{\circ}$}

\newcommand{\brjpsiphsp} { $(3.29\pm 0.20  \pm 0.26  )\times 10^{-5}$}
\newcommand{\brjpsirho}  { $(7.90\pm 0.19  \pm 0.49  )\times 10^{-5}$}
\newcommand{\brjpsiomega}{ $(2.08\pm 0.30  \pm 0.14  )\times 10^{-4}$}
\newcommand{\brjpsirhop} { $(3.28\pm 0.55  \pm 0.44  )\times 10^{-6}$}
\newcommand{\brpsipphspI}{ $(5.13\pm 1.23  \pm 0.64  )\times 10^{-6}$}
\newcommand{\brpsiprhoI} { $(1.02\pm 0.11  \pm 0.24  )\times 10^{-5}$}
\newcommand{\brpsipphspII}{$(5.13\pm 1.14  \pm 0.62  )\times 10^{-6}$}
\newcommand{\brpsiprhoII} {$(5.69\pm 1.28  \pm 2.36  )\times 10^{-6}$}
\newcommand{\brpsiprhopdgI}  {$(1.9^{+1.7}_{-1.2})\times 10^{-5}$}
\newcommand{\brpsiprhopdgII} {$(19.^{+17.}_{-12.})\times 10^{-6}$}
\newcommand{\brjpsirhopdg}  {$(10.5\pm 1.8 )\times 10^{-5}$ }
\newcommand{\brjpsiomegapdg}{$(1.82\pm 0.21)\times 10^{-4}$}
\newcommand{\ratiophspI} {$15.6\pm 3.9  \pm 2.3 $} %
\newcommand{\ratiophspII}{$15.6\pm 3.6  \pm 2.3 $}
\newcommand{\ratiorhoI}  {$12.9\pm 1.4  \pm 3.1 $}
\newcommand{\ratiorhoII} {$7.2 \pm 1.6  \pm 3.0 $}

\newcommand{\brjpsiphsptot} {$(1.36 \pm 0.02  \pm 0.08 )\times 10^{-4}$}
\newcommand{\brpsipphsptot} {$(1.51 \pm 0.14  \pm 0.23 )\times 10^{-5}$}
\newcommand{\ratiototal} {$11.1 \pm 1.0  \pm 1.8 $}

\begin{document}


\title{\boldmath Study of $J/\psi$ and $\psi(3686)$ decays to $\pi^+\pi^-\eta^\prime$}

\author{
\begin{small}
\begin{center}
M.~Ablikim$^{1}$, M.~N.~Achasov$^{9,d}$, S. ~Ahmed$^{14}$, M.~Albrecht$^{4}$, A.~Amoroso$^{53A,53C}$, F.~F.~An$^{1}$, Q.~An$^{50,40}$, J.~Z.~Bai$^{1}$, Y.~Bai$^{39}$, O.~Bakina$^{24}$
% ...
\\
\vspace{0.2cm}
(BESIII Collaboration)\\
\vspace{0.2cm} {\it
$^{1}$ Institute of High Energy Physics, Beijing 100049, People's Republic of China\\
% ...
$^{4}$ Bochum Ruhr-University, D-44780 Bochum, Germany\\
% ...
$^{9}$ G.I. Budker Institute of Nuclear Physics SB RAS (BINP), Novosibirsk 630090, Russia\\
% ...
$^{14}$ Helmholtz Institute Mainz, Johann-Joachim-Becher-Weg 45, D-55099 Mainz, Germany\\
% ...
$^{24}$ Joint Institute for Nuclear Research, 141980 Dubna, Moscow region, Russia\\
% ...
$^{39}$ Southeast University, Nanjing 211100, People's Republic of China\\
$^{40}$ State Key Laboratory of Particle Detection and Electronics, Beijing 100049, Hefei 230026, People's Republic of China\\
% ...
$^{50}$ University of Science and Technology of China, Hefei 230026, People's Republic of China\\
% ...
$^{53}$ (A)University of Turin, I-10125, Turin, Italy; (B)University of Eastern Piedmont, I-15121, Alessandria, Italy; (C)INFN, I-10125, Turin, Italy\\
% ...
$^{d}$ Also at the Novosibirsk State University, Novosibirsk, 630090, Russia\\
% ...
}\end{center}

\vspace{0.4cm}
\end{small}}

\begin{abstract}
Using the data samples of $1.31\times 10^9$ $J/\psi$ events and $4.48\times 10^8$ $\psip$ events collected with the BESIII
detector, partial wave analyses on the decays $J/\psi$ and $\psip \to \pi^+\pi^-\eta^\prime$ are performed with a relativistic
covariant tensor amplitude approach. The dominant contribution is found to be $J/\psi$ and $\psip$ decays to $\rho\eta^\prime$. In the $J/\psi$ decay, the branching fraction ${\cal B}(J/\psi\to \rho\eta^\prime)$ is determined to be $(7.90\pm0.19(\mathrm{stat})\pm0.49(\mathrm{sys}))\times 10^{-5}$.  Two solutions are found in the $\psip$ decay, and the corresponding branching fraction ${\cal B}(\psip\to \rho\eta^\prime)$ is $(1.02\pm0.11(\mathrm{stat})\pm0.24(\mathrm{sys}))\times 10^{-5}$ for the case of destructive interference, and $(5.69\pm1.28(\mathrm{stat})\pm2.36(\mathrm{sys}))\times 10^{-6}$ for constructive interference. As a consequence, the ratios of branching fractions between $\psip$ and $J/\psi$ decays to $\rho\eta^\prime$ are calculated to be
$(12.9\pm1.4(\mathrm{stat})\pm3.1(\mathrm{sys}))$\% and $(7.2\pm1.6(\mathrm{stat})\pm3.0(\mathrm{sys}))$\%, respectively.
We also determine the inclusive branching fractions of $\jpsi$ and $\psip$ decays to $\pi^+\pi^-\eta^\prime$ to be $(1.36\pm0.02(\mathrm{stat})\pm0.08(\mathrm{sys}))\times 10^{-4}$ and $(1.51\pm0.14(\mathrm{stat})\pm 0.23(\mathrm{sys}))\times 10^{-5}$,
respectively.

\end{abstract}


\maketitle

\section{\boldmath INTRODUCTION}
The decays of $\psi$ mesons ($\psi$ denotes both the $\jpsi$ and $\psip$ charmonium states throughout the text) provide an excellent laboratory in which to explore the various hadronic properties and strong interaction dynamics in a nonperturbative regime~\cite{BESIIIBOOK}. In particular, the decay
$\psi \to \rho\eta^\prime$ is an isospin symmetry breaking process. The measurement of its branching fraction will shed
light on the isospin breaking effects in $\psi \to VP$ (where $V$ and $P$ represent vector and pseudoscalar mesons, respectively) decays~\cite{qianwang}, and can be also used to calculate
the associated electromagnetic form factors~\cite{Brodsky}, which are used to test quantum chromodynamics (QCD) inspired models of
mesonic wave functions. In the framework of perturbative QCD (pQCD), the partial width for the $\psi$ decays into
an exclusive hadronic final state is expected to be proportional to the square of the $c\bar{c}$ wave function overlap at the origin,
which is well determined from the leptonic width~\cite{trule}.
Thus the ratio of branching fractions of $\psip$ and $\jpsi$ decays to any specific final state $h$ is expected to be
\begin{eqnarray}
  \label{Q12}
  Q_h = \frac{{\cal B}(\psip\to h)}{{\cal B}(J/\psi\to h)} \simeq &&  \frac{{\cal B}(\psip\to e^+e^-)}
  {{\cal B}(J/\psi\to e^+e^-)} \nonumber\\
   \simeq&& 12.7\%,
\end{eqnarray}
which is the well known ``12\% rule". Although the rule works well for some decay modes, it fails spectacularly
in the $\psi$ decays to $VP$ \cite{Brodsky,vpsuppress2} such as $\psi \to \rho\pi$~\cite{rhopipuzzle}.
A precise measurement of the branching fraction for $\psi$ decays to $\rho\eta^\prime$ also provides a good opportunity to test the ``12\% rule". The current world average branching fraction of the $\jpsirho$ decay is $\mathcal{B}(\jpsi\to\rho\etap)=(1.05\pm0.18)\times 10^{-4}$, according to the particle data group (PDG)~\cite{pdg}. This value has not been updated for about 30 years since the measurements by the DM2~\cite{DM2} and MARK-III~\cite{MARKIII} experiments.
 For $\psip\to \rho\eta^\prime$, the only available branching fraction, ${\cal B}(\psip \to \rho\eta^\prime) = (1.9^{+1.7}_{-1.2})\times 10^{-5}$, was measured by the BESII experiment~\cite{BESpsip}.

In this paper, using the samples of $1.31\times 10^9$ $J/\psi$ events~\cite{jpsinumber} and $4.48\times 10^8$ $\psip$ events~\cite{psipnumber09,psipnumber12} accumulated with the Beijing Spectrometer
III (BESIII) detector~\cite{bes} operating at the Beijing Electron-Positron Collider II (BEPCII)~\cite{bepc}, a partial
wave analysis (PWA) of the decay $\psi \to \pi^+\pi^-\eta^\prime$ is performed. The intermediate contribution is found to be dominated by $\psi \to \rho\eta^\prime$, and the corresponding branching fractions are determined.

\section{\boldmath{Detector and Monte Carlo Simulation}}

BEPCII is a double-ring electron-positron collider operating in the center-of-mass energy ($\sqrt s$) range from 2.0 to 4.6~GeV.
The design peak luminosity of $10^{33}$~cm$^{-2}$s$^{-1}$ was reached in 2016, with a beam current of 0.93~A at $\sqrt s$ = 3.773~GeV.
The cylindrical core of the BESIII detector consists of a helium-based main drift chamber (MDC), a plastic scintillator time-of-flight
(TOF) system, a CsI(Tl) electromagnetic calorimeter (EMC), a superconducting solenoidal magnet providing
a 1.0~T (0.9~T in 2012) magnetic field, and a muon system (MUC) made of resistive plate chambers in the iron flux return yoke of the magnet.
The acceptances for charged particles and photons are 93\% and 92\% of 4$\pi$, respectively. The charged particle momentum resolution is 0.5\% at 1~GeV/$c$, and the barrel (endcap) photon energy resolution is 2.5\% (5.0\%) at 1~GeV.

The optimization of the event selection and the estimation of physics backgrounds are performed using Monte Carlo (MC) simulated samples. The {\sc geant4}-based~\cite{genant4} simulation software {\sc boost}~\cite{boost} includes the geometric and material description of
the BESIII detector, the detector response and digitization models, as well as a record of the detector running conditions
and performance. The production of the $\psi$ resonance is simulated by the MC event generator {\sc kkmc}~\cite{kkmc}.
The known decay modes are generated by {\sc evtgen}~\cite{evtgen1,evtgen2} by setting branching ratios to be the world average values~\cite{pdg12},
and by {\sc lundcharm}~\cite{lundcharm} for the remaining unknown decays. A MC generated event is mixed with a randomly triggered event recorded in data taking to consider the possible background contamination, such as beam-related backgrounds and cosmic rays, as well as the electronic noise and hot wires. The analysis is performed in the framework of the BESIII
offline software system which takes care of the detector calibration, event reconstruction and data storage.

\section{\boldmath{Event Selection}}

Charged tracks in an event are reconstructed from hits in the MDC. We select tracks within $\pm$10~cm of the interaction
point in the beam direction and within 1~cm in the plane perpendicular to the beam. The tracks must have a polar angle $\theta$ satisfying
$|\cos\theta|<0.93$. The time-of-flight and energy loss ($dE/dx$) information are combined to evaluate particle identification (PID)
probabilities for the $\pi$, $K$, and $e$ hypotheses; each track is assigned to the particle type corresponding to the hypothesis
with the highest confidence level. Electromagnetic showers are reconstructed from clusters of energy deposited in the EMC. The energy
deposited in nearby TOF counters is included to improve the reconstruction efficiency and energy resolution. The photon candidate
showers must have a minimum energy of 25~MeV in the barrel region ($|\cos\theta|<0.80$) or 50~MeV in the end cap region ($0.86<|\cos\theta|<0.92$).
To suppress showers from charged particles, a photon must be separated by at least $10^\circ$ from the nearest charged track.
Timing information from the EMC for the photon candidates must be in coincidence with collision events, \textit{i.e.}, $0\leq t \leq 700$ ns, to suppress electronic noise and energy deposits unrelated to the event.

 The cascade decay of interest is $\psi \to \pi^{+} \pi^{-} \etap$, $\etap \to \eta \pi^{+} \pi^{-}$, and $\eta \to \gamma \gamma$. Candidate events are required to have four charged tracks with zero net charge and at least two photon candidates. A four-constraint (4C)
kinematic fit imposing overall energy-momentum conservation is performed to the $\gamma\gamma\pi^+\pi^-\pi^+\pi^-$ hypothesis, and the events with $\chi^2_{\mathrm{4C}}<40$ are retained. For events with more
than two photon candidates, the combination with the least $\chi^2_{\mathrm{4C}}$ is selected. Further selection criteria are based on the four-momenta from the kinematic fit. The $\eta$ candidate is reconstructed with the selected $\gamma\gamma$ pair, and must have an invariant mass in the range (0.525, 0.565)~GeV/$c^2$.

\begin{figure}[tb]
\centering
   \includegraphics[width=0.4\textwidth]{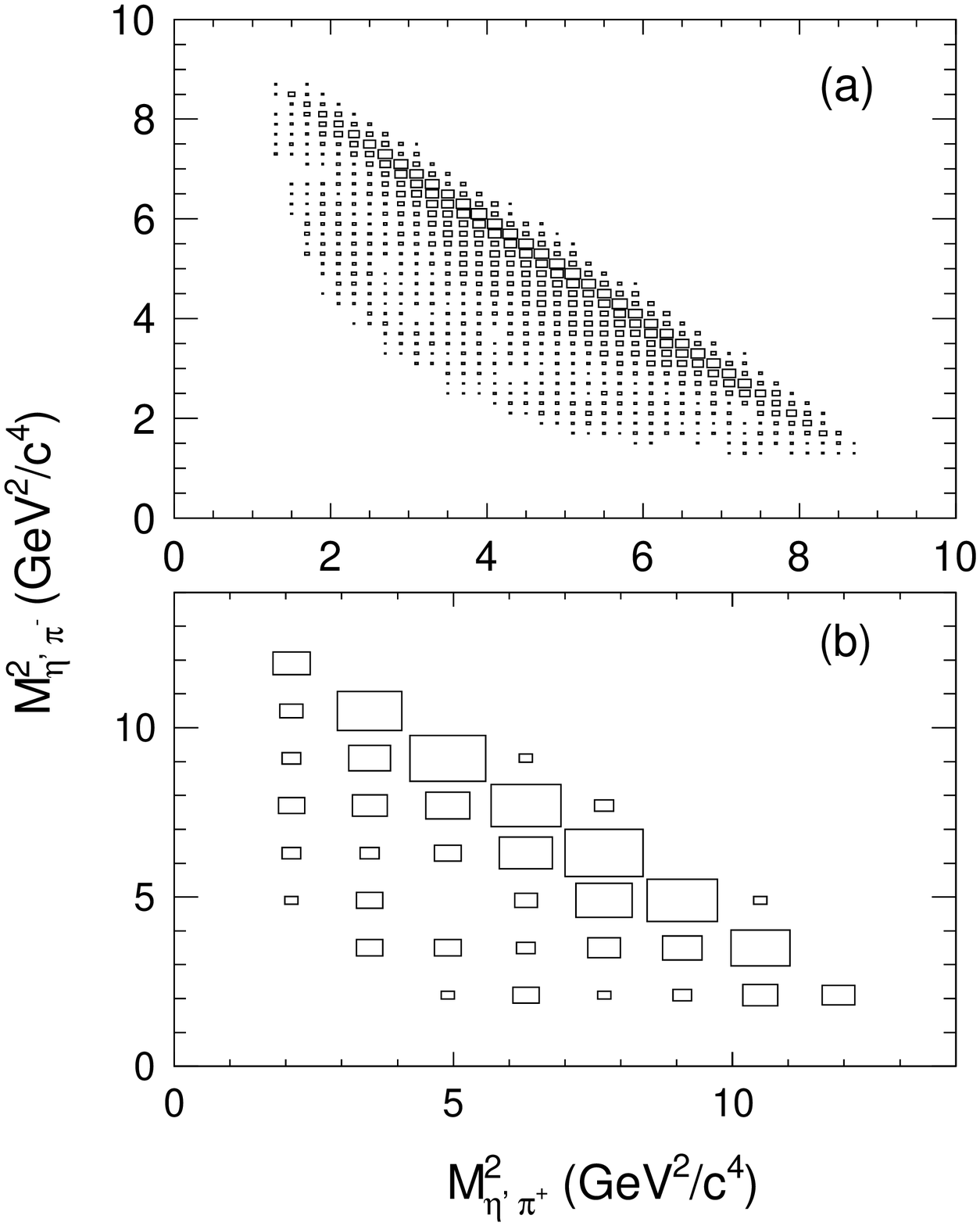}
  \caption{Dalitz plots for (a) $J/\psi\to\pi^+\pi^-\eta^\prime$ and (b) $\psi(3686) \to \pi^+\pi^-\eta^\prime$ with events in
  the $\eta^\prime$ signal region.}
\label{dalitz}
\end{figure}

After the above requirements, the $\eta^\prime$ candidate is reconstructed from the $\eta\pi^{+}\pi^{-}$ combination whose invariant mass $M_{\eta \pi^{+} \pi^{-}}$ is closest to the $\etap$ nominal mass~\cite{pdg}. The $\eta^\prime$ signal
region is defined as $0.935<M_{\eta\pi^+\pi^-}< 0.975$~GeV/$c^2$. A total of 7016 and 313 candidate events for $\jpsi$ and $\psip$ data, respectively, survive the event selection
criteria. The corresponding Dalitz plots of $M^{2}_{\etap \pi^{+}}$ versus $M^{2}_{\etap \pi^{-}}$ are depicted in Fig.~\ref{dalitz}, where bands along the diagonal, corresponding to the decay $\psi \to \rho \etap$, are clearly visible.

%

\section{\boldmath{BACKGROUND ANALYSIS}}\label{sectionbkg}

The inclusive MC samples of $1.23\times 10^9$ $J/\psi$ and $5.06\times 10^8$ $\psip$ events are used to study potential backgrounds. According to the MC study, the backgrounds in the $\jpsi$ decay can be categorized into two classes. The class I backgrounds are dominated by the decays $J/\psi\to 2(\pi^+\pi^-)\eta$ with $\eta\to \gamma\gamma$, and $J/\psi\to \gamma\pi^+\pi^-\eta$ with $\eta\to \gamma\pi^+\pi^-$, which do not include an $\etap$ intermediate state. The class II background mainly arises from the decay $\jpsidimu$, with $\mu^{\pm}$ misidentified as a $\pi^{\pm}$, which produces a peak in the distribution of $M_{\eta \pi^{+}\pi^{-}}$. In the $\psip$ decay, only class I backgrounds appear, which are dominated by $\psip \to 2(\pi^+\pi^-)\eta$ and $\psip \to \eta\jpsi$ with $\jpsi \to 2(\pi^+\pi^-)$ and $\eta\to \gamma\gamma$, and the class II background is negligible.

In the analysis, the class I
backgrounds can be estimated using the events in $\eta^\prime$ sideband regions, which are defined as $0.85<M_{\eta\pi^+\pi^-}<0.90$~GeV/$c^2$
and $1.00<M_{\eta\pi^+\pi^-}<1.05$~GeV/$c^2$.
The class II background in $\jpsi$ decay, which is dominated by the decay $\jpsidimu$, is estimated with the MC simulation. Considering the consistency of the branching fraction ${\cal B}(J/\psi\to
e^+e^-P)$ ($P$ represents $\eta$ and $\eta^\prime$ mesons) between the experimental measurements~\cite{chuxk} and
the theoretical calculations~\cite{jpsidimu}, the MC sample for $\jpsidimu$ is generated according to the amplitude in Ref.~\cite{jpsidimu}. Using the same selection criteria and taking the branching fraction ${\cal B}(J/\psi\to \mu^+\mu^-\eta^\prime) = (1.31\pm0.04)\times 10^{-5}$ quoted in Ref.~\cite{jpsidimu}, $661\pm23$ events are expected for this peaking background.

The background from the continuum process $e^{+}e^{-} \to \pi^{+} \pi^{-}\etap$ under the $\psi$ peak is studied using the off-resonance samples of 153.8 pb$^{-1}$ taken at $\sqrt{s} = 3.08$~GeV and 48.8 pb$^{-1}$ taken at $\sqrt{s} = 3.65$~GeV. With the same selection criteria, $81\pm10$ and $5\pm2$
  events survive from the off-resonance samples taken at $\sqrt{s} = 3.08$~GeV and $\sqrt{s} = 3.65$~GeV, respectively. These background events have the same final state as and are indistinguishable from signal. Therefore, the contributions from the continuum process are subtracted directly from the obtained signal yields.

\section{\boldmath{Fitting $M_{\eta\pi^+\pi^-}$ spectrum} }\label{sectionfit}

\begin{figure}[tb]
\centering
    \includegraphics[width=0.45\textwidth]{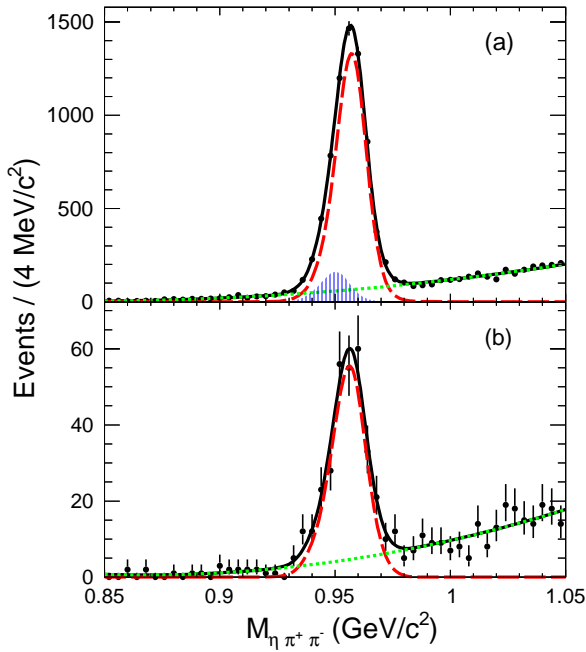}
  \caption{(color online). Distribution of $M_{\eta\pi^+\pi^-}$ for (a) $J/\psi$ and (b) $\psi(3686)$ decays. The red dashed line is the signal
  MC shape convolved with a Gaussian, the green dotted line is the class I backgrounds described by a $2^\text{nd}$ Chebychev function, the hatched
  area is the class II background, dominated by $J/\psi \to \mu^{+} \mu^{-} \eta^{\prime}$, described by the MC simulation, the black solid line is the overall fit result, and the
  dots with error bars are the data.}
\label{etapmass}
\end{figure}

After applying all selection criteria, the numbers of candidate events for $J/\psi$ and $\psip$ decays to $\pi^{+} \pi^{-} \etap$ are obtained to be $5730\pm86$ and
$264\pm18$, respectively, by performing an unbinned maximum likelihood fit to the $M_{\eta\pi^+\pi^-}$ spectra. In the fit, the signal shape is modeled by the MC simulation convolved with
a Gaussian function with free parameters to account for the data-MC difference in detector resolution. The shape of the class I backgrounds
is described by a $2^\text{nd}$ order Chebychev function, and the class II background is modeled with the MC simulation of $\jpsidimu$ decay with the number of expected events described in Sec.~\ref{sectionbkg}. Figure~\ref{etapmass} shows the fitted $M_{\eta\pi^+\pi^-}$ spectra for the $J/\psi$ and $\psip$ data.

%

\section{\boldmath{Partial Wave Analysis}}\label{pwamethod}
\subsection{\boldmath{Analysis method}}
\label{sectionpwa}

A PWA is performed on the selected $\psi \to \pi^+\pi^-\eta^\prime$ candidate events. The quasi two-body decay
amplitudes in the sequential decay process $\psi \to X\eta^\prime$, $X\to \pi^+\pi^-$ are constructed using the
covariant tensor amplitudes described in Ref.~\cite{zoubs}. The general form for the decay amplitude $A$ of a vector meson $\psi$ with spin projection $n$ is
\begin{eqnarray}
  \label{Amp}
  A = \psi_\mu(n) A^\mu = \psi_\mu(n)\sum_a \Lambda_a U_a^\mu,
\end{eqnarray}
where $\psi_\mu(n)$ is the polarization vector of the $\psi$ meson, $U_a^\mu$ is the $a$-th partial-wave amplitude with a coupling
strength $\Lambda_a$, which is a complex number. The specific expressions are introduced in Ref.~\cite{zoubs}.

The $a$-th partial amplitude $U_a$ includes a Blatt-Weisskopf barrier factor~\cite{zoubs}, which is used
to damp the divergent tail due to the momentum factor of $p^l$ in the decay $A\to B + C$, where the $p$ and $l$ are the momentum of particle
$B$ in the rest system of particle $A$ and the relative orbital angular momentum between particle $B$ and $C$, respectively. From
a study in Ref.~\cite{barrier}, the radius of the centrifugal barrier is taken to be 0.7~fm in this analysis.

The intermediate state $X$ is parameterized by a Breit-Wigner (BW) propagator. In this analysis, two different BW
propagators are used. One is described with a constant width
\begin{eqnarray}
  \label{BW}
  \mathrm{BW} = \frac{1}{m^2 - s - im\Gamma},
\end{eqnarray}
where $s$ is the invariant mass-squared of $\pi^+\pi^-$, and $m$ and $\Gamma$ are the mass and width of the intermediate state.
The other BW propagator is parameterized using the Gounaris-Sakurai (GS) model~\cite{GS1,GS2}, which is appropriate for states like the $\rho$ meson and its excited states,
\begin{eqnarray}
  \label{BW-GS}
  \mathrm{BW^{GS}} = \frac{1 + d(m)\Gamma/m}{m^2 - s + f(s, m, \Gamma) - im\Gamma(s, m, \Gamma)},
\end{eqnarray}
with
\begin{eqnarray}
  \Gamma(s, m, \Gamma) =&& \Gamma \frac{s}{m^2}\left(\frac{\beta_\pi(s)}{\beta_\pi(m^2)}\right)^3, \nonumber\\
                  d(m) =&& \frac{3}{\pi} \frac{m^2_\pi}{k^2(m^2)} \ln\left(\frac{m + 2k(m^2)}{2m_\pi}\right) + \frac{m}{2\pi k(m^2)}  \nonumber\\
                        &&- \frac{m^2_\pi m}{\pi k^3(m^2)}, \\
       f(s, m, \Gamma) =&& \frac{\Gamma m^2}{k^3(m^2)}\left[k^2(s)(h(s) - h(m^2))\right. \nonumber \\
                        &&+\left. (m^2 - s)k^2(m^2)h'(m^2)\right], \nonumber
\end{eqnarray}
where
\begin{eqnarray}
    \beta_\pi(s) =&& \sqrt{1 - 4m_\pi^2/s}, \nonumber\\
             k(s)=&& \frac{1}{2} \sqrt{s} \beta_\pi(s), \\
             h(s)=&& \frac{2}{\pi} \frac{k(s)}{\sqrt{s}}\ln\left(\frac{\sqrt{s} + 2k(s)}{2m_\pi}\right), \nonumber
\end{eqnarray}
and $h^\prime(s)$ is the derivative of $h(s)$.

The complex coefficients of the amplitudes and the resonance parameters are determined by an unbinned maximum likelihood fit.
The likelihood fit is used to calculate the probability that a hypothesized probability density function (PDF) can produce the data set under consideration.
The probability to observe the $i$-th event characterized by the measurement $\xi_i$, \textit{i.e.}, the measured four-momenta of the particles
in the final state, is the differential observed cross section normalized to unity
\begin{eqnarray}
  \label{PDF0}
  P(\xi_i, \alpha) = \frac{\omega(\xi_i, \alpha) \epsilon(\xi_i)}{\int d\xi_i \omega(\xi_i, \alpha) \epsilon(\xi_i)},
\end{eqnarray}
where $\omega(\xi_i, \alpha) \equiv (\frac{d\sigma}{d\Phi})_i$ is the differential observed cross section, $\alpha$ is a set of unknown parameters
to be determined in the fit, $d\Phi$ is the standard element of phase space, $\epsilon(\xi_i)$ is the detection efficiency,
and $\int d\xi_i \omega(\xi_i, \alpha) \epsilon(\xi_i) \equiv \sigma^\prime$ is the total observed cross section. The full differential observed cross section is
\begin{eqnarray}
  \label{DifCS}
  \frac{d\sigma}{d\Phi} = \frac{1}{2}\sum_{\mu=1}^{2} A^{\mu}A^{*\mu},
\end{eqnarray}
where $\mu = 1,2$ means the direction of the $x$- and $y$-axis, respectively, and $A$ is the total amplitude for all possible resonances.

The joint PDF for observing $N$ events in the data sample is
\begin{eqnarray}
  \label{PDF1}
   \mathcal{L} = \prod\limits_{i=1}^N P(\xi_i, \alpha) = \prod\limits_{i=1}^N \frac{\omega(\xi_i, \alpha) \epsilon(\xi_i)}
   {\int d\xi_i \omega(\xi_i, \alpha) \epsilon(\xi_i)}.
\end{eqnarray}

MINUIT~\cite{Minuit,qzhzhao} is used to optimize the fitted parameters to achieve the maximum likelihood value. Technically, rather
than maximizing $\mathcal L$, $\mathcal S = -\ln\mathcal L$ is minimized, \textit{i.e.},
\begin{eqnarray}
  \label{lnL}
  \mathcal{S} &=& -\ln\mathcal{L} \nonumber\\
              &=& -\sum_{i=1}^N \ln\frac{\omega(\xi_i, \alpha)}{\int d\xi_i \omega(\xi_i, \alpha) \epsilon(\xi_i)} -
                  \sum_{i=1}^N \ln \epsilon(\xi_i).
\end{eqnarray}
For a given data set, the second term is a constant and has no impact on the relative changes of the $\mathcal S$ values.

We take the detector resolution into account by convoluting the probability $P(x)$ with a Gaussian function $G_\sigma(x)$. The
variable $x$ represents the invariant mass of $\pi^+\pi^-$ ($M_{\pi^+\pi^-}$), and $P(x)$ is the same as $ P(\xi_i, \alpha)$. The redefined probability $u(x)$
is
\begin{eqnarray}
  \label{Conv}
  u(x) = (P \otimes G_{\sigma})(x) = \int G_{\sigma}(x - y) P(y)dy.
\end{eqnarray}
We use an approximate method~\cite{res_cleoc,chdfu}
to calculate Eq.~(\ref{Conv}), \textit{i.e.}, the effect of smearing is considered by numerically convoluting the detector resolution with
the probability at each point when performing the fit, at 11 points from $-5\sigma$ to 5$\sigma$. Hence the convolution is turned
into a sum,
\begin{eqnarray}
  \label{Convsum}
  u(x)  = &&\sum^{5\sigma}_{m=-5\sigma}g_m P(x - m)\Delta m, \nonumber \\
  g_{m} = &&\frac{1}{T} G_\sigma(m), \\
      T = &&\sum^{5\sigma}_{m=-5\sigma}G_\sigma(m)\Delta m,~
 \Delta m = \sigma,\nonumber
\end{eqnarray}
where $g_m$ is the value of the Gaussian function normalized to unity at the point $m$, $T$ is the sum value of the Gaussian function for 11 points. In this analysis, the resolution $\sigma$ of $M_{\pi^{+}\pi^{-}}$ is 3~MeV/$c^{2}$, as determined from MC simulations.

The background (not including the continuum process here) contribution to the log-likelihood is estimated with the weighted events in the $\etap$ sideband regions for the class I backgrounds and with MC simulated $\jpsidimu$ events (in the $\jpsi$ decay only) for the class II
background, and is subtracted from the log-likelihood value of data in the $\eta^\prime$ signal region, \textit{i.e.},
\begin{eqnarray}
  \label{eqlikelihood}
  {\cal S} = && -(\ln{\cal L}_\text{data} - \ln{\cal L}_\text{bkg}).
\end{eqnarray}

The number of fitted events $N_X$ for a given intermediate state $X$,
is obtained by
\begin{eqnarray}
  \label{xnumber}
  N_X = f_X N^\prime = \frac{\sigma_X}{\sigma^\prime} N^\prime,
\end{eqnarray}
where $N^\prime$ is the number of selected events after background subtraction, and $f_X$ is the ratio between the observed cross section $\sigma_{X}$ for the intermediate state $X$ and the total observed cross section $\sigma'$. Both $\sigma_{X}$ and $\sigma'$ are calculated with the MC simulation approach according to the fitted amplitudes. A signal MC sample of $N_{\rm{gen}}$ events is generated with a uniform distribution in phase space. These events
are subjected to the same selection criteria and $N_{\rm{acc}}$ events are accepted. The observed cross sections of the overall process and a given state $X$ are computed as
\begin{eqnarray}
  \label{totalsigma}
  \sigma' \rightarrow \frac{1}{N_\text{acc}} \sum_{i}^{N_\text{acc}}\left(\frac{d\sigma}{d\Phi}\right)_{i},
\end{eqnarray}
 and
\begin{eqnarray}
  \label{xsigma}
  \sigma_{X} = \frac{1}{N_\text{acc}}
  \sum_{i}^{N_\text{acc}} \left(\left.\frac{d\sigma}{d\Phi}\right|_{X}\right)_{i} ,
\end{eqnarray}
respectively, where $\frac{d\sigma}{d\Phi}|_{X}$ denotes the differential observed cross section for the process with the intermediate state $X$.

The branching fraction of $\psi \to X\eta^\prime$ is evaluated by
\begin{eqnarray}\label{branchingfraction}
  {\mathcal B}(\psi \to X\eta^{\prime}) = \frac{N_X}{N_{\psi} \, \varepsilon_{X}  \, {\mathcal B}},
\end{eqnarray}
where $N_\psi$ is the total number of $\psi$ events, the detection efficiency
$\varepsilon_X$ is obtained using the weighted MC sample,
\begin{eqnarray}\label{calefficiency}
 \varepsilon_{X} = \frac{ \sum_{i}^{N_\text{acc}} \left(\left.\frac{d\sigma}{d\Phi}\right|_{X}\right)_{i} } { \sum_{i}^{N_\text{gen}} \left(\left.\frac{d\sigma}{d\Phi}\right|_{X}\right)_{i}}
\end{eqnarray}
and ${\mathcal B}$ = ${\mathcal B}({X\to \pi^+\pi^-}){\mathcal B}({\eta^\prime\to \pi^+\pi^-\eta})
{\mathcal B}({\eta\to \gamma\gamma})$ is the product of the decay branching fractions in the subsequent decay chain. All branching fractions are quoted from the world average values~\cite{pdg}.

In order to estimate the statistical uncertainty of the branching fraction ${\mathcal B}(\psi \to X\eta^{\prime})$ associated with the statistical uncertainties of the fit parameters, we repeat the calculation several hundred times by randomly varying the fit parameters according to the error matrix~\cite{qzhzhao}. Then we fit the resulting distribution with a Gaussian function, and take the fitted width as the statistical uncertainty.

\subsection{\boldmath{PWA of $\psi\to \pi^+\pi^-\eta^\prime$ decay}}\label{pwajpsi}

Due to spin-parity and angular momentum conservation, in the $\psi \to X\eta^\prime$, $X\to \pi^+\pi^-$ process, $X$ must have
$J^{PC}$ of $1^{--}$, $3^{--}$, $\cdots$. In this analysis, only the intermediate states $X$ with $J^{PC} = 1^{--}$ are considered,
since the higher spin states would encounter a power suppression due to the large orbital angular momentum. The intermediate states $\rho$, $\omega$ and other possible excited $\rho$ states listed in the PDG~\cite{pdg} as well as
a non-resonant (NR) contribution are included in the fit. The contribution from the combination of broad vector mesons with higher masses like excited $\rho$ mesons is expected. Since we are not able to describe the contribution of all possible mesons individually, we include it in the model using the NR amplitude constructed by a three-body phase space with a $J^{PC} = 1^{--}$ angular distribution for the $\pi^{+}\pi^{-}$ system.
However, only the components with a statistical significance larger than 5$\sigma$ are kept
as the basic solution, where the statistical significance of a state is evaluated by considering the change in the likelihood values and the
numbers of free parameters in the fit with and without the state included.

\begin{figure}[tb]
  \centering
     \includegraphics[width=0.49\textwidth]{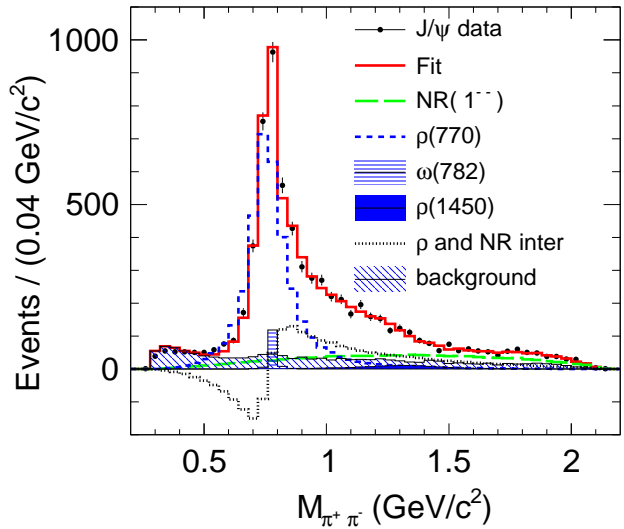}
  \caption{(color online). Comparisons of the distributions of $M_{\pi^+\pi^-}$ between data and PWA fit projections for the decay $J/\psi\to \pi^+\pi^-\eta^\prime$.}
  \label{pwa:jpsi}
\end{figure}

\begin{figure}[htbp]
  \centering
   \includegraphics[width=0.45\textwidth]{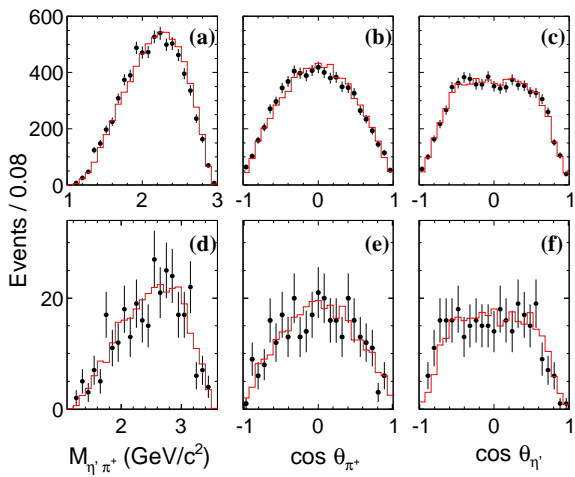}
  \caption{(color online). Comparisons between data and PWA fit projections for the decay $J/\psi \to \pi^+\pi^-\eta^\prime$, shown in the first row,
  and for the decay $\psi(3686) \to \pi^+\pi^-\eta^\prime$, shown in the bottom row. The left is for the distributions of $M_{\eta^\prime\pi^+}$, and the middle and the right are for the distributions of $\cos\theta_{\pi^{+}}$ and $\cos\theta_{\eta}$. The dots with error bars are data, and the red solid line is
  the PWA fit projection.}
  \label{pwa:jpsi2}
\end{figure}

In the decay $J/\psi \to \pi^+\pi^-\eta^\prime$, the mass and width of the $\omega$ meson are fixed to the world average values~\cite{pdg}.
The basic fitted solution is found to contain four components, namely the $\rho$, $\omega$, $\rho(1450)$ intermediate states as well as the NR contribution. The PWA fit projections on $M_{\pi^+\pi^-}$, the invariant mass of $\etap \pi^+$ ($M_{\etap\pi^+}$), as well as the polar angle of $\etap$ ($\pi^{+}$) in the $\jpsi$ ($\pi^+\pi^-$) helicity frame $\cos\theta_{\etap}$ ($\cos\theta_{\pi^{+}}$) are
shown in Fig.~\ref{pwa:jpsi} and Fig.~\ref{pwa:jpsi2} (first row). The $M_{\pi^+\pi^-}$ distributions for the individual components are also shown in Fig.~\ref{pwa:jpsi}. The statistical significances are larger than 30$\sigma$ for $\rho$ component, and equal to 12.5$\sigma$, 10.7$\sigma$ and 8.0$\sigma$ for $\rho(1450)$, $\omega$ and the NR components, respectively.
The mass and width returned by the fit are ($766\pm2$)~MeV/$c^{2}$ and ($142\pm5$)~MeV for the $\rho$ meson, and ($1369\pm38$)~MeV/$c^{2}$ and ($386\pm70$)~MeV for the $\rho(1450)$ meson, respectively. These are in good agreement with the previous measurements~\cite{pdg, rho1450} within uncertainties. The phase angles for the $\rho(1450)$, $\omega$ and NR components relative to the $\rho$ component are ($203.6 \pm 11.9$)\deg, ($100.3\pm5.3$)\deg~and ($-269.7\pm1.4$)\deg, respectively. We also try to add the cascade decay $\psi \to X^{\pm} \pi^{\mp}$ with decay $\X^{\pm}\to \etap \pi^{\pm}$ in the fit, where $X$ can be the $a_{2}(1320)$ or other possible states in the PDG~\cite{pdg}. But all these processes are found to have the statistical significances less than $5\sigma$.

\begin{figure}[tb]
\centering
   \includegraphics[width=0.45\textwidth]{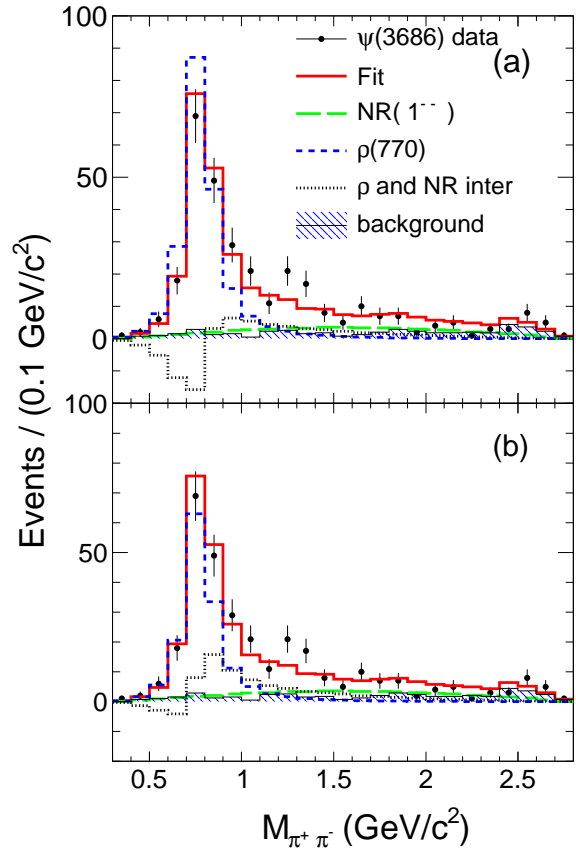}
  \caption{(color online). Comparisons of the distributions of $M_{\pi^+\pi^-}$ between data and PWA fit projections for the decay $\psi(3686) \to
  \pi^+\pi^-\eta^\prime$ with (a) destructive and (b) constructive interferences.}
\label{pwa:psip}
\end{figure}

The same fit procedure is performed to the data sample for $\psipphsp$. The basic solution includes a $\rho$ component interfering with NR component due to the low statistics. In the fit, the mass and width of the $\rho$ meson are fixed to the world average values~\cite{pdg}. Two solutions with the same fit quality are found, corresponding to the case of destructive and constructive interference between the two components with a relative phase angle $(120.3\pm16.6)^{\circ}$ and $(45.6\pm17.5)^{\circ}$, respectively. A dedicated study on the mathematics for the multiple solutions is discussed in Ref.~\cite{multisolution}. The $\rho$ and NR components are observed with statistical significances
of 20$\sigma$ and 15.1$\sigma$, respectively. The PWA fit projections on $M_{\etap\pi^+}$ , $\cos\theta_{\pi^+}$ and $\cos\theta_{\eta^\prime}$, are shown in Fig.~\ref{pwa:jpsi2} (bottom row). The $M_{\pi^{+}\pi^{-}}$ distribution and the fit curve as well as the individual components are shown in Fig.~\ref{pwa:psip} for the case of destructive and constructive interference, individually.

\subsection{\boldmath{PWA of off-resonance data}}\label{pwaoffres}
 A similar PWA fit is performed on the accepted data sample at $\sqrt{s} = 3.08$~GeV, which yields the numbers of events $58\pm11$ and $11\pm3$ for the $\rho$ and NR components, with statistical significances of $11.1\sigma$ and $6.6\sigma$, respectively. The contributions from the intermediates $\omega$ and $\rho(1450)$ are neglected because of the low statistical significances of 0.8$\sigma$
  and 1.5$\sigma$, respectively. Due to the low statistics at $\sqrt s = 3.65$~GeV, we assume the dominant contribution is from the $\rho$ component. Taking into account the integrated luminosities of the off-resonance sample and $\psi$ data, as well as the central energy dependence of the production cross section (proportional to $1/s$), we determine the normalized number of events for $e^{+}e^{-}\to \rho\etap$ to be $145\pm28$ and $68\pm 27$ for the $\jpsi$ and $\psip$ data samples, respectively, and $28\pm8$ for the NR process in the $\jpsi$ data sample.

\section{\boldmath SYSTEMATIC UNCERTAINTIES}\label{Sys_err}
The sources of systematic uncertainty and their contributions to the uncertainty in the measurements of branching fractions for $\psi \to
X\eta^\prime$ and inclusive $\psi \to \pi^{+} \pi^{-} \etap$ decays are described below.

The systematic uncertainties can be divided into two categories. The first category is from the event selection,
including the uncertainties on the photon detection efficiency, MDC tracking efficiency, trigger efficiency, PID efficiency, the kinematic fit, the $\eta$ and $\etap$ mass window requirements, the cited branching
fractions, and the number of $\psi$ events. The second category includes uncertainties associated with the PWA fit procedure.

\begin{table}[tb]
\begin{center}
 \caption{Relative systematic uncertainties from the event selection (in percent).}
  \label{err_Evt}
  \setlength{\extrarowheight}{1.0ex}
  \renewcommand{\arraystretch}{1.0}
  \vspace{0.2cm}
  \begin{tabular}{p{3.8cm}m{2.3cm}<{\centering}m{2.0cm}<{\centering}}
  \hline\hline
  Source                                    &$J/\psi\to \pi^+\pi^-\eta^\prime$  &$\psip\to \pi^+\pi^-\eta^\prime$  \\ \hline
  Photon detection                          &1.2                                &1.2              \\
  MDC tracking                              &4.0                                &4.0            \\
  Trigger efficiency                        &negligible                         &negligible \\
  PID                                       &4.0                                &4.0            \\
  Kinematic fit                             &0.3                                &1.0            \\
  $\eta$ mass window                        &0.5                                &0.7             \\
  $\etap$ mass window                       &0.6                                &1.1             \\
  Cited branching fractions                 &1.7                                &1.7             \\
  N$_{\mathrm{\psi}}$                       &0.5                                &0.6              \\\hline
  Total                                     &6.1                                &6.3             \\
  \hline\hline
  \end{tabular}
  \vspace{-0.2cm}
  \end{center}
\end{table}

The systematic uncertainty due to the photon detection efficiency is studied using a control sample of $J/\psi \to \pi^+\pi^-\pi^0$,
and determined to be 0.5\% per photon in the EMC barrel and 1.5\% per photon in the EMC endcap. Thus, the uncertainty associated with the two  reconstructed photons is 1.2\% (0.6\% per photon) by weighting the uncertainties according to the polar angle distribution of the two photons from real data. The uncertainty due to
the charged tracking efficiency has been investigated with control samples of $\jpsi \rightarrow \rho \pi$ and $\jpsi \rightarrow p \bar{p} \pi^{+} \pi^{-}$~\cite{psiptrack}, and a difference of 1\% per
track between data and MC simulation is considered as the systematic uncertainty. The uncertainty arising from the trigger efficiency is negligible
according to the studies in Ref.~\cite{triggereff}. The uncertainty due to PID efficiency has been studied with control samples
of $J/\psi\to \pi^+\pi^-\pi^0$ and $\psip\to \gamma\chi_{cJ}$, $\chi_{cJ}\to \pi^+\pi^-\pi^+\pi^-$, and the difference in PID efficiencies between the data and MC simulation is determined to be 4.0\% (1.0\% per track).  This is taken as the systematic uncertainty.

A systematic uncertainty associated with the kinematic fit occurs due to the inconsistency of track-helix parameters between the data and MC simulation. Following the procedure described in Ref.~\cite{4cfit}, we use $J/\psi\to \pi^+\pi^-\pi^0$ and $\psip \to \pi^{+} \pi^{-} \jpsi~(\jpsi \to \mu^{+} \mu^{-})$ decays as the control
sample to determine the correction factors of the pull distributions of the track-helix parameters for the $J/\psi$ and $\psip$ decays, respectively. We estimate the detection efficiencies using MC samples with and without the corrected helix parameters for the charged tracks, and the resulting differences in the detection efficiencies, 0.3\% for the $J/\psi$ sample and 1.0\% for the $\psip$ sample, are assigned as the systematic
uncertainties associated with the kinematic fit.

The systematic uncertainty arising from the $\eta$ ($\etap$) mass window requirement is evaluated by changing the mass window from (0.525, 0.565) GeV/$c^{2}$ to (0.52, 0.57) GeV/$c^{2}$ (from (0.935, 0.975) GeV/$c^{2}$ to (0.93, 0.98) GeV/$c^{2}$). The difference in the branching fractions of the inclusive decay $\psi \to \pi^{+} \pi^{-} \etap$ is taken as the systematic uncertainty associated with the $\eta$ ($\etap$) mass window requirement, which is 0.5 (0.6)\% for $\jpsi$ decay and 0.7 (1.1)\% for $\psip$ decay, respectively.

The uncertainties associated with the branching fractions of $\eta^\prime\to \pi^+\pi^-\eta$ and $\eta\to \gamma\gamma$ are taken from the world average values~\cite{pdg}. The number of $\psi$ events used in the analysis is $N_{J/\psi} = (1310.6 \pm 7.0)\times 10^6$~\cite{jpsinumber} and $N_{\psip} =
(448.1\pm2.9)\times 10^6$~\cite{psipnumber09,psipnumber12}, which is determined by counting the hadronic events. The uncertainty is 0.5\% for the $\jpsi$ decay and 0.6\% for the $\psip$ decay, respectively.

All of the above systematic uncertainties, summarized in Table~\ref{err_Evt}, are in common for all branching fraction measurements in this analysis. The total systematic uncertainty, which is the quadratic sum of the individual values assuming all the sources of uncertainty are independent, is 6.1\% for the $\jpsi$ decay and 6.3\% for the $\psip$ decay, respectively.

The category of uncertainties associated with the PWA fit procedure affect the branching fraction measurement of $\psi \to X \etap$. The sources and the corresponding uncertainties are discussed in detail below.
\begin{enumerate}[(i)]

  \item The uncertainty due to the barrier factor is estimated by varying the radius of the centrifugal barrier~\cite{barrier}
  from 0.7~fm to 0.6~fm. The change of the signal yields is taken as the systematic uncertainty.

  \item The uncertainty associated with the BW parametrization is evaluated by the changes of the signal yields when replacing the GS BW for the
  $\rho$ and $\rho(1450)$ mesons with a constant-width BW.

  \item In the nominal PWA fit, the detector resolution on $M_{\pi^{+}\pi^{-}}$ is parameterized using a constant value of 3~MeV/$c^2$. An alternative fit is performed
  with a mass-dependent detector resolution, which is obtained from the MC simulations of the decay $\psi \to X \etap$, $X \to \pi^{+}\pi^{-}$, generated with different masses for the $X$ ($1^{--}$) meson. The changes in the resulting branching fractions are taken as the systematic uncertainties.

  \item In the nominal PWA fit, the mass and width of the $\omega$ meson are fixed to the world average values~\cite{pdg} in the $J/\psi$ decay, and those of the $\rho$ meson are fixed in the $\psip$ decay. To evaluate the uncertainty associated with the mass and width of the $\omega$ ($\rho$) meson, we repeat the fit by changing its mass and width by one standard deviation according to the world average values~\cite{pdg}. The resulting changes on the branching fractions are taken as the systematic uncertainties.

  \item To estimate the uncertainty from extra resonances, alternative fits are performed by adding the $\rho(1700)$ meson and the cascade decay process $\jpsi \to a_{2}(1320)^{\pm} \pi^{\mp} \to \pi^{+} \pi^{-}\etap$ for the $\jpsi$ data sample, and the $\omega$, $\rho(1450)$ and $\rho(1700)$ mesons for the $\psip$ data sample, into the baseline configuration individually. The largest changes in the resulting branching fractions are assigned as the systematic uncertainties.

  \item In the PWA fit, the effect on the likelihood fit from class I backgrounds is estimated using the events in the $\etap$ sideband regions. We repeat the fit with an alternative sideband regions $(0.85, 0.91) \cup (0.99, 1.04)$~GeV/$c^2$ for the class I backgrounds, and the resulting change in the measured branching fractions is regarded as the systematic uncertainty. The uncertainty related to the class II background
  $J/\psi\to \mu^+\mu^-\eta^\prime$ in the PWA fit of $J/\psi\to \pi^+\pi^-\eta^\prime$ is evaluated by varying the number
  of expected events by one standard deviation according to the uncertainty in the theoretically predicted branching fraction in Ref.~\cite{jpsidimu}.
  The change of the resulting branching fractions is taken as the systematic uncertainty. The contributions from the continuum processes are estimated with the off-resonance data samples, and subtracted from the signal yields directly. The corresponding uncertainties are propagated to the measured branching fractions. The systematic uncertainties from backgrounds of class I, class II, and the continuum process are summed in quadrature.
\end{enumerate}
%

The total systematic uncertainty in the measured branching fraction for the decay $\psi$ $\to X\eta^\prime$ is obtained by summing the individual systematic uncertainties in quadrature, as summarized in Table~\ref{error:tot}.
\begin{table*}
  \begin{center}
  \caption{Relative systematic uncertainties for the branching fraction measurement of the decay $\psi \to X\eta^\prime$
  (in percent).}
  \label{error:tot}
  \setlength{\extrarowheight}{1.0ex}
  \renewcommand{\arraystretch}{1.0}
  \vspace{0.2cm}
  \begin{tabular}{p{4.0cm}|m{1.4cm}<{\centering}m{1.4cm}<{\centering}m{1.4cm}<{\centering}m{1.4cm}<{\centering}|m{1.4cm}<{\centering}m{1.4cm}<{\centering}|m{1.4cm}<{\centering}m{1.0cm}<{\centering}}
  \hline\hline
  \multirow{3}{*}{Source} &\multicolumn{4}{c|}{$J/\psi$ decay}          &\multicolumn{4}{c}{$\psip$ decay}  \\ \cline{2-9}
  &  \multicolumn{4}{l|}{}   &\multicolumn{2}{c}{Solution I}  &\multicolumn{2}{c}{Solution II}  \\ \cline{6-9}
                       &NR  &$\rho$ &$\omega$ &$\rho$(1450)  &NR &$\rho$ &NR &$\rho$ \\ \hline
  Event selection      &6.1  &6.1  &6.1  &6.1     &6.3  &6.3  &6.3  &6.3  \\
  Barrier factor       &3.0  &0.5  &0.1  &4.9     &7.1  &1.0  &6.8  &2.7  \\
  Breit-Wigner formula &0.7  &0.4  &0.4  &1.7     &4.8  &10.2 &4.4  &4.3  \\
  Detector resolution  &0.0  &0.1  &1.6  &0.1     &0.0  &0.0  &0.1  &0.1  \\
  Resonance parameters &0.1  &0.3  &0.2  &0.1     &0.6  &0.2  &0.5  &0.2  \\
  Extra resonances     &3.3  &0.5  &1.0  &9.4     &5.4  &7.4  &5.4  &22.6 \\
  Background           &2.6  &0.8  &1.2  &5.0     &3.8  &19.0 &3.1  &33.9  \\

  \hline
  Total                &8.0  &6.2  &6.5  &13.3    &12.5 &23.7 &12.0 &41.5  \\
  \hline\hline
  \end{tabular}
  \vspace{0.2cm}
  \end{center}
\end{table*}

The systematic uncertainties in the measurement of the branching fraction for the inclusive decay $\psi \to \pi^+\pi^-\eta^\prime$ are coming from the event selection (listed in Table~\ref{err_Evt}), signal shape, background estimation, and PWA. In the nominal fit
to the $M_{\eta\pi^+\pi^-}$ distribution, the signal PDF is described by the MC shape convolved with a Gaussian function. An alternative
fit is performed by modeling the signal shape with the MC simulation only, and the resultant change in yields is considered as the systematic uncertainty.
The uncertainties due to the backgrounds of class I, class II and continuum processes are evaluated by changing the order of the Chebychev polynomial
function from $2^\text{nd}$ to $3^\text{rd}$, varying the expected number of events for the decay $J/\psi\to \mu^+\mu^-\eta^\prime$  and continuum processes by one standard
deviation, respectively. The systematic uncertainty is determined to be 0.6\% and 13.9\% for $J/\psi$ and $\psip$ decays, respectively. The event selection efficiency for the inclusive $\psi \to \pi^+\pi^-\eta^\prime$ decay is obtained with MC simulations according to the nominal PWA solution. An alternative MC sample is simulated by changing the fit parameters by one standard deviation. The resulting difference in the detection efficiencies is taken as the systematic uncertainty due to the PWA.

The total systematic uncertainty on the inclusive branching fraction for $\psi \to \pi^+\pi^-\eta^\prime$
 is the quadratic sum of the individual contributions, as summarized in Table~\ref{sysfitetaptotal}.
\begin{table}[tb]
\begin{center}
  \caption{Relative systematic uncertainties for the inclusive branching fraction of $\psi \to \pi^+\pi^-\eta^\prime$ decay (in percent).}
  \label{sysfitetaptotal}
  \setlength{\extrarowheight}{1.0ex}
  \renewcommand{\arraystretch}{1.0}
  \vspace{0.2cm}
  \begin{tabular}{p{3.0cm}m{2.5cm}<{\centering}m{2.0cm}<{\centering}}
  \hline\hline
  Source                   &$\jpsi$  &$\psip$ \\
  \hline
  Event selection          &6.1                                      &6.3            \\
  Signal shape             &0.3                                      &1.1            \\
  Background shape         &0.6                                      &13.9            \\
  PWA                      &0.7                                      &2.3            \\\hline
  Total                    &6.2                                      &15.5            \\
  \hline\hline
  \end{tabular}
  \vspace{-0.2cm}
  \end{center}
\end{table}
%

\section{\boldmath{RESULTS AND  DISCUSSION}}

The signal yields of $\psi$ and off-resonance data samples, detection efficiencies and branching fractions are summarized in Table~\ref{BR}.
The ratios of branching fractions between $\psip$ and $J/\psi$ decays to the same final states are listed in Table~\ref{ratios},
where the correlated systematic uncertainties between the $\jpsi$ and $\psip$ decays, arising from the photon efficiency, MDC tracking, PID, trigger
efficiency, kinematic fit, $\eta$ and $\etap$ mass window requirements and the cited branching fractions, are canceled.

\begin{table*}[htbp]
\begin{center}
  \caption{The signal yields for the $\psi$~$(N_{0})$ and off-resonance data ($N_{c}$) samples, the detection efficiency ($\varepsilon$) for each component, as well as the measured branching fractions ($\mathcal{B}$) in this work and values from PDG~\cite{pdg}, where the
  first uncertainties are statistical and the second are systematic. Here Inc represents inclusive decay and ``-" means ignoring the effect from the continuum process.}
  \label{BR}
  \setlength{\extrarowheight}{1.0ex}
  \renewcommand{\arraystretch}{1.0}
  \vspace{0.2cm}
  \begin {tabular}{p{3.5cm}m{2.0cm}<{\centering}m{2.5cm}<{\centering}m{2.0cm}<{\centering}m{3.7cm}<{\centering}m{3.7cm}<{\centering}}
  \hline\hline
  Channel& $N_{0}$ & $N_{c}$   &$\varepsilon(\%)$ &${\mathcal B}$  &PDG           \\ \hline
  $\jpsirho$    &$3621\pm83$              &$145\pm28$    &$20.0$   &\brjpsirho   &  \brjpsirhopdg \\
  $\jpsiomega$  &$137\pm20$               &-     &$19.6$   &\brjpsiomega &  \brjpsiomegapdg\\
  $J/\psi\to \rho(1450)\eta^\prime$, &\multirow{2}{*}{$119\pm20$} &\multirow{2}{*}{-} & \multirow{2}{*}{$16.5$} & \multirow{2}{*}{\brjpsirhop} & \\
  $\rho(1450)\to \pi^+\pi^-$         &                  &                            &                         &                              &  \\
  $\jpsiphsp_{(\text{NR})}$  &$1214\pm72$  &$28\pm8$     &$16.4$   &\brjpsiphsp  &  \\
  $\jpsiphsp_{(\text{Inc})}$ &$5730\pm86$  &$203\pm25$    &$18.5$   &\brjpsiphsptot & \\
  \hline\hline
  \multicolumn{5}{c}{Solution I}\\
  $\psiprho$                  &$211\pm16$ &$68\pm27$     &$18.7$   &\brpsiprhoI  &\brpsiprhopdgI  \\
  $\psipphsp_{(\text{NR})}$   &$54\pm13$  &-      &$14.0$   &\brpsipphspI &                \\
  \hline
  \multicolumn{5}{c}{Solution II}\\
  $\psiprho$                  &$148\pm18$ &$68\pm27$     &$18.7$   &\brpsiprhoII &\brpsiprhopdgII \\
  $\psipphsp_{(\text{NR})}$   &$54\pm12$  &-     &$14.0$   &\brpsipphspII&                \\
  \hline
  $\psipphsp_{(\text{Inc})}$  &$264\pm18$ &$68\pm27$     &$17.2$   &\brpsipphsptot & \\
  \hline\hline
  \end{tabular}
  \vspace{-0.2cm}
\end{center}
\end{table*}

\begin{table*}[htbp]
\begin{center}\normalsize
  \caption{The ratios of branching fractions between $\psip$ and $J/\psi$ decay to $\rho\eta^\prime$, NR and inclusive
  decays (\%). The first uncertainties are statistical and the second systematic.}
  \label{ratios}
  \setlength{\extrarowheight}{0.8ex}
  \renewcommand{\arraystretch}{0.8}
  \vspace{0.2cm}
  \begin{tabular}{p{5.0cm}m{4.7cm}<{\centering}m{4.7cm}<{\centering}}
  \hline\hline
                                                         &Solution I                                &Solution II  \\
                                                         \hline
  \multirow{2}{*}{$\frac{{\cal B}(\psip \to \pi^{+}\pi^{-}\eta^{\prime})_{(\text{NR})}}{{\cal B}(\jpsi \to \pi^{+}\pi^{-}\eta^{\prime})_{(\text{NR})}}$} &\multirow{2}{*}{\ratiophspI}&\multirow{2}{*}{\ratiophspII}  \\
  &&\\
  \multirow{2}{*}{$\frac{{\cal B}(\psip \to \rho\eta^{\prime})}{{\cal B}(J/\psi\to \rho\eta^{\prime})}$} &\multirow{2}{*}{\ratiorhoI}&\multirow{2}{*}{\ratiorhoII}  \\
  &&\\
  \hline
  \multirow{2}{*}{$\frac{{\cal B}(\psip \to \pi^{+}\pi^{-}\eta^{\prime})_{(\text{Inc})}}{{\cal B}(\jpsi \to \pi^{+}\pi^{-}\eta^{\prime})_{(\text{Inc})}}$}
  &\multicolumn{2}{c}{\multirow{2}{*}{\ratiototal}} \\
  &&\\
  \hline\hline
  \end{tabular}
  \vspace{-0.2cm}
\end{center}
\end{table*}

With the yields of the continuum processes from the off-resonance data samples, we can estimate the branching fraction of $\psi \to \rho \etap$ based on some hypotheses. Compared with the measurement, we can test these hypotheses.

Assuming that the decay $\psi\to \rho\eta^\prime$ is a pure electromagnetic process, which is caused by one virtual photon
exchange, from factorization we have the following relation according to Ref.~\cite{daijp12},
\begin{eqnarray}
  \label{emequation}
  \frac{\sigma(e^+e^-\to \gamma^*\to \rho\eta^\prime)}{\sigma(e^+e^-\to \psi\to \rho\eta^\prime)}\approx
  \frac{\sigma(e^+e^-\to \gamma^*\to \mu^+\mu^-)}{\sigma(e^+e^-\to \psi\to \mu^+\mu^-)}.
\end{eqnarray}
    At the $\psi$ peak, for the specific final state $h$ we can have $\sigma$($e^{+}e^{-} \to \psi \to h$) = $\mathcal{B}$($\psi\to h$) $N_{\psi} /\mathcal{L_{\psi}}$ by neglecting the interference between $e^{+}e^{-} \to \gamma^{*} \to h$ and $e^{+}e^{-} \to \psi \to h$,
where $\mathcal{L}_{\psi}$ is the corresponding integrated luminosity.  Thus one can get
\begin{eqnarray}
  \label{emequation123}
   \mathcal{B}(\psi\to \rho\eta^\prime) \simeq && \nonumber \\
     \mathcal{B}(\psi\to &&\mu^+\mu^-)  \frac{\sigma(e^+e^-\to \gamma^*\to \rho\eta^\prime)}{\sigma(e^+e^-\to \gamma^*\to \mu^+\mu^-)} .
\end{eqnarray}

Using the observed $e^{+}e^{-}\to \rho \etap$ signal events $N_\text{obs}$ and the integrated luminosity $\mathcal{L}$ of the off-resonance data sample, the detection efficiency $\epsilon$ from MC simulation and the initial state radiative (ISR) correction factor $f$ (1.1 for $\sqrt{s} = 3.08$~GeV and 1.3 for $\sqrt{s} = 3.65$~GeV, respectively), the cross section of $e^{+}e^{-}\to \gamma^{*} \to \rho \etap$ is calculated to be $10.2\pm 1.9$~pb at $\sqrt{s}= 3.08$~GeV and $2.5\pm 1.0$~pb at $\sqrt{s} = 3.65$~GeV, respectively, according to the formula $N_\text{obs}/(\mathcal{L}\, \varepsilon\,  f\, \mathcal{B})$, where $\mathcal{B}$ is the product branching fraction in the cascade decay $\mathcal{B}$ = $\mathcal{B}(\rho\to \pi^+\pi^-)\, \mathcal{B}(\eta^\prime\to \eta\pi^+\pi^-)\, \mathcal{B}(\eta\to \gamma\gamma)$ quoted from the world average value~\cite{pdg}. Taking into account the cross section of $e^{+}e^{-}\to \gamma^{*} \to \rho \etap$ measured above, and of $e^{+}e^{-}\to \gamma^{*} \to \mu^{+}\mu^{-}$ in Ref.~\cite{crosssectioneetomumu} (9.05~nb at $\sqrt s = 3.08$~GeV and 6.4~nb at $\sqrt s = 3.65$~GeV), as well as the world average decay branching fraction of $\mathcal{B}(\psi\to\mu^{+}\mu^{-})$ to the Eq.~(\ref{emequation123}), we obtain the estimated branching fractions of $\mathcal{B}$($\jpsi\to\rho\etap$)$_{\mathrm{ES}}$ = ($6.72\pm1.25$) $\times 10^{-5}$ and $\mathcal{B}$($\psip\to\rho\etap$)$_{\mathrm{ES}}$ = ($3.09\pm1.23$) $\times 10^{-6}$, respectively.

Based on the above calculation, we also obtain the ratio of branching fractions for the decay $\psi \to \rho \etap$ between this measurement ($\mathcal{B}$($\psi \to \rho \etap$)$_{\mathrm{MS}}$) and the estimation from the off-resonance data ($\mathcal{B}$($\psi \to \rho \etap$)$_{\mathrm{ES}}$), as listed in Table~\ref{qedresult}, where the systematic uncertainties for the ratio are from the number of $\psi$ events, the luminosity of off-resonance data sample (1.0\%), the ISR factor (1.0\%) and the cited branching fraction of $\jpsi \to \mu^{+}\mu^{-}$ (0.6\%) or $\psip \to \mu^{+}\mu^{-}$ (11.4\%). From the table, we find that the branching fractions of $\psi \to \rho \etap$ between the measurement from the $\psi$ resonant data and the estimation from off-resonance data sample are consistent within 1$\sigma$ for the $\jpsi$ decay and the $\psip$ decay with the constructive solution, while they are within 2$\sigma$ for the $\psip$ decay with the destructive solution. The hypotheses used in the theoretical estimation are acceptable based on our current data.

\begin{table*}[!bt]
  \begin{center}\normalsize
  \caption{The ratio of branching fractions of $\psi\to \rho \etap$ between our measurement (MS) and estimation (ES).}
  \label{qedresult}
  \setlength{\extrarowheight}{0.8ex}
  \renewcommand{\arraystretch}{0.8}
  \vspace{0.2cm}
  \begin{tabular}{p{5.0cm}m{4.7cm}<{\centering}m{4.7cm}<{\centering} }
  \hline\hline
  & Solution I & Solution II \\
  \hline
  \multirow{2}{*}{$\frac{\mathcal{B}(\psip \to \rho \etap)_{\text{MS}}}{\mathcal{B}(\psip \to \rho \etap)_{\text{ES}}}$} & \multirow{2}{*}{$3.31 \pm 1.37 \pm 0.60$}  & \multirow{2}{*}{$1.84 \pm 0.85 \pm 0.33$} \\
  &\\
  \multirow{2}{*}{$\frac{\mathcal{B}(\jpsi \to \rho \etap)_{\text{MS}}}{\mathcal{B}(\jpsi \to \rho \etap)_{\text{ES}}}$} & \multicolumn{2}{c}{\multirow{2}{*}{$1.18 \pm 0.22 \pm 0.02$}}\\
  &\\
  \hline\hline
  \end{tabular}
  \vspace{0.2cm}
  \end{center}
\end{table*}

\section{\boldmath SUMMARY}\label{summary}

In summary, using samples of $1.31\times 10^9$ $J/\psi$ events and $4.48\times 10^8$ $\psip$ events collected with the BESIII
detector, partial wave analyses of $J/\psi\to \pi^+\pi^-\eta^\prime$ and $\psip \to \pi^+\pi^-\eta^\prime$ decays are
performed.
For the $\jpsi$ decay, besides the dominant contribution from $\jpsirho$ decay, contributions from $\jpsiomega$, $\jpsi \to \rho(1450) \etap$ and NR $\jpsiphsp$ are found to be necessary in the PWA. In the $\psip$ decay, due to low statistics, the PWA indicates that only two components, $\psip \to \rho\etap$ and NR $\psipphsp$  are sufficient to describe the data. The same fit quality is obtained with either destructive or constructive interference between the two components. Using the PWA results,
we obtain the branching fractions for the processes with different intermediate components and the inclusive decay $\psi \to \pi^{+} \pi^{-} \etap$, as listed in Table~\ref{BR}.

With these measurements, we obtain the ratio of branching fractions between $\psip$ and $\jpsi$ decays to $\rho \etap$ final states, ~(\ratiorhoI)\%~and ~(\ratiorhoII)\%~for the case of destructive and constructive interference in the $\psip$ data, respectively, as listed in Table~\ref{ratios}. These do not obviously violate the ``12\%" rule within one standard deviation. We also assume that the isospin violating decay $\psi \to \rho \etap$ occurs via a pure electromagnetic process and estimate its branching fraction with off-resonance data samples at $\sqrt{s} = 3.08$ and 3.65~GeV. From Table~\ref{qedresult}, we find the estimated branching fractions of $\psi \to \rho \etap$ are consistent with those from the data at the resonant $\psi$ peak.

  \section*{\boldmath ACKNOWLEDGEMENTS}
  The BESIII collaboration thanks the staff of BEPCII and the IHEP computing center for their strong support. This work is supported in part by National Key Basic Research Program of China under Contract No. 2015CB856700; National Natural Science Foundation of China (NSFC) under Contracts Nos. 11235011, 11335008, 11425524, 11505111, 11625523, 11635010; the Chinese Academy of Sciences (CAS) Large-Scale Scientific Facility Program; the CAS Center for Excellence in Particle Physics (CCEPP); Joint Large-Scale Scientific Facility Funds of the NSFC and CAS under Contracts Nos. U1332201, U1532257, U1532258; CAS under Contracts Nos. KJCX2-YW-N29, KJCX2-YW-N45, QYZDJ-SSW-SLH003; 100 Talents Program of CAS; National 1000 Talents Program of China; INPAC and Shanghai Key Laboratory for Particle Physics and Cosmology; German Research Foundation DFG under Contracts Nos. Collaborative Research Center CRC 1044, FOR 2359; Istituto Nazionale di Fisica Nucleare, Italy; Koninklijke Nederlandse Akademie van Wetenschappen (KNAW) under Contract No. 530-4CDP03; Ministry of Development of Turkey under Contract No. DPT2006K-120470; National Science and Technology fund; The Swedish Research Council; U. S. Department of Energy under Contracts Nos. DE-FG02-05ER41374, DE-SC-0010118, DE-SC-0010504, DE-SC-0012069; University of Groningen (RuG) and the Helmholtzzentrum fuer Schwerionenforschung GmbH (GSI), Darmstadt; WCU Program of National Research Foundation of Korea under Contract No. R32-2008-000-10155-0.

%

\end{document}